\documentclass[a4paper,11pt]{article}
\pdfoutput=1 % if your are submitting a  (i.e. if you have
\usepackage{jheppub} % for details on the use of the package, please
\usepackage[T1]{fontenc} % if needed

\usepackage{lineno}

\def\Ecm {\ensuremath{\rm E_{\rm c.m.}}}
\def\Eb {\ensuremath{\rm E_{\rm beam}}}

\def\epem {\ensuremath{e^+ e^-}}
\def\pipi {\ensuremath{\pi^+\pi^-}}

\def\mevcc {\ensuremath{\rm MeV/c^2}}
\def\piz {\ensuremath{\pi^0}}

\def\Ks {\ensuremath{K_{S}^{0}}}
\def\Kl {\ensuremath{K_{L}^{0}}}
\def\Kn {\ensuremath{K^{0}}}

\title{ \boldmath
Neutral kaon mass measurement with the CMD-3 detector at VEPP-2000
}

\author[a,b,1]{E.P.~Solodov\note{Corresponding author: solodov@inp.nsk.su}}
\author[d]{N.A.~Petrov}
\author[a,b]{R.R.~Akhmetshin}
\author[a,b]{A.N.~Amirkhanov}
\author[a,b]{A.V.~Anisenkov}
\author[a,b]{V.M.~Aulchenko}
\author[a]{N.S.~Bashtovoy}
\author[a]{D.E.~Berkaev}
\author[a,b]{A.E.~Bondar}
\author[a]{A.V.~Bragin}
\author[a,b]{D.A.~Epifanov}
\author[a,b,c]{L.B.~Epshteyn}
\author[a,b]{A.L.~Erofeev}
\author[a,b]{G.V.~Fedotovich}
\author[a,c]{A.O.~Gorkovenko}
\author[a,b]{A.A.~Grebenuk}
\author[a,b]{S.S.~Gribanov}
\author[a,b,c]{D.N.~Grigoriev}
\author[a,b]{F.V.~Ignatov}
\author[a]{D.R.~Ivanov}
\author[a,b]{V.L.~Ivanov}
\author[a]{S.V.~Karpov}
\author[a,b]{V.F.~Kazanin}
\author[a,b]{I.A.~Koop}
\author[a,b]{A.A.~Korobov}
\author[a,c]{A.N.~Kozyrev}
\author[a,b]{P.P.~Krokovny}
\author[a,b]{A.S.~Kuzmin}
\author[a,b]{I.B.~Logashenko}
\author[a,b]{P.A.~Lukin}
\author[a,b]{K.Yu.~Mikhailov}
\author[a,b]{I.V.~Obraztsov}
\author[a]{A.V.~Otboev}
\author[a]{Yu.N.~Pestov}
\author[a,b]{E.A.~Perevedentsev}
\author[a,b]{A.S.~Popov}
\author[a,b]{Yu.A.~Rogovsky}
\author[a]{A.A.~Ruban}
\author[a]{\fbox{N.M.~Ryskulov}}
\author[a,b]{A.E.~Ryzhenenkov}
\author[a,b]{A.V.~Semenov}
\author[a]{A.I.~Senchenko}
\author[a]{Yu.M.~Shatunov}
\author[a,b]{V.E.~Shebalin}
\author[a,b]{D.N.~Shemyakin}
\author[a,b]{B.A.~Shwartz}
\author[e]{D.B.~Shwartz}
\author[a]{M.V.~Timoshenko}
\author[a]{V.M.~Titov}
\author[a,b]{A.A.~Talyshev}
\author[a]{S.S.~Tolmachev}
\author[a]{A.I.~Vorobiov}
\author[a]{I.M.~Zemlyansky}
\author[a]{D.S.~Zhadan}
\author[a,b]{Yu.V.~Yudin}

\affiliation[a]{Budker Institute of Nuclear Physics, SB RAS, 
Novosibirsk, 630090, Russia}
\affiliation[b]{Novosibirsk State University, Novosibirsk, 630090, Russia}
\affiliation[c]{Novosibirsk State Technical University, Novosibirsk, 630092, Russia}
\affiliation[d]{Institute for Nuclear Research, RAS, Moscow, 117312, Russia}
\affiliation[e]{P-cure Ltd, Shilat, 7318800, Israel}

\vspace{0.7cm}
\abstract{
  Using more than 600 thousand
  $K_{S}^{0}\to\pi^{+}\pi^{-}$ decays from the $e^{+}e^{-}\to\phi(1020)\to K_{S}^{0}K_{L}^{0}$ 
  reaction, the neutral kaon mass has been measured with the CMD-3 detector at the VEPP-2000 collider.
Using the beam energies control by the back-scattering laser light photons, and a calibration to the
world average $\phi$-meson mass, the neutral kaon mass is determind to be 
m($K^0$) = 497.587 $\pm$ 0.004(stat.) $\pm$ 0.008(syst.) $\pm$
0.009(calibr.) \mevcc, where the first uncertainty is statistical, the second is systematic,
and the third is the uncertainty from the energy calibration.
}

\begin{document} 
%Version 11.0  ~~\today

\maketitle
\flushbottom

%\linenumbers

\baselineskip=17pt
\section{ \boldmath Introduction}
\hspace*{\parindent}
The neutral kaon mass was first measured in the 1960s with the optical spark
counters and the bubble chambers with about 0.3$\div$0.5 \mevcc~
accuracy~\cite{pdg} using neutral kaon beams or kaons from the
proton-anti-proton annihilation.  A real breakthrough happened in the 1980s
when the neutral kaons became available at the \epem~ colliders
with a precision measurement of the beam energies by the
resonance depolarization method~\cite{rdm}.
A clean kinematics in the $\epem\to\phi(1020)\to\Ks\Kl$ reaction
allows one to develop a method to determine a kaon momentum by
measuring an opening angle between pions from the $\Ks\to\pipi$ decay.
In the specific case when the pions in the kaon rest frame 
move perpendicular to the kaon momentum direction, the opening angle has
a minimum value. 
In this paper we call it  "edge angle'', $\psi_c$, and the kaon mass is
calculated as:  
\begin{equation}
{\rm m}(\Ks) = \sqrt{{\rm E}^2_{\Ks}{\rm
    sin^2}\frac{\psi_c}{2}+4{\rm m}^2_{\pi}{\rm cos^2}\frac{\psi_c}{2}}.
\label{kmass0}
\end{equation}
The equation includes only the kaon energy  (${\rm E}_{\Ks}$  well measured), the pion
mass value (${\rm m}_\pi$ very precisely known), and the edge angle. 
This method was used for the first time in the  collider
experiments by CMD collaboration~\cite{barkov85,barkov87}, and allowed to
get the kaon mass with a few tens of keV/c$^2$. 
The precision of these measurements was  limited by the low
collider luminosity and by beam energy drift during the data taking.

A  more general kinematical expression, suggested in
ref.~\cite{Zaitsev}, allows one to use pions with any decay angles to
calculate the kaon mass.
  This approach has been used
  by the CMD-2 Collaboration, reaching an overall  systematic
  uncertainty of 24 keV/c$^2$.
The result was reported and appeared in a conference
proceedings~\cite{Zaitsev}, but had not been published in a regular peer-reviewed  journal.
Later, high-statistics measurements of the neutral  kaon mass were published by several experimental
groups~\cite{NA48,KLOE,KLEO}. These measurements are in tension with the first measurements
by the CMD Collaboration, with a scale factor of 1.2~\cite{pdg}. Further clarification
would therefore be desirable.

In this paper we present a new measurement of the neutral kaon mass with
the CMD-3 detector at the VEPP-2000 collider~\cite{vepp}. The neutral
kaons are produced in the dedicated energy scan of the $\phi(1020)$ resonance with
the integrated luminosity of about 14 pb$^{-1}$, collected at 15 energy
points. The machine is equipped with the
the Back-Scattering-Laser-Light system~\cite{laser}, which provides
a continuous beam energy monitoring during the
data taking, essential for the measurement. The high luminosity of the
VEPP-2000 collider allows us to collect and analyze a large number 
of the neutral kaons.
The edge angle approach is used for the presented measurement. 
The detector has three-four times better angular resolution for the
charged particles than that of CMD-2, therefore much smaller resolution corrections are
required.  
The large statistics allow us to perform many cross checks
and reduce  systematic uncertainties. The world-average mass value~\cite{pdg} for
the $\phi(1020)$ resonance  is used for the absolute energy calibration.

\section{ \boldmath CMD-3 detector}
\hspace*{\parindent}
The general-purpose detector CMD-3 has been described in 
detail elsewhere~\cite{sndcmd3}. Its tracking system consists of a 
cylindrical drift chamber (DC)~\cite{dc}  placed inside a thin 
(0.2~X$_0$) superconducting solenoid with a field of 1.3~T.

The tracking system allows one to detect charged particle tracks with a minimum polar 
angle of about 0.5 radians with respect to the beam axis (about 90\% of 4$\pi$).
The DC information allows us to measure the charged particle  momenta
with 1.5-4.5\% accuracy in the
100-1000 MeV/c range, and provides the measurement of the polar  and azimuth
 angles with 15--20 mrad and 3.5-8.0 mrad accuracies, respectively. An
 amplitude information from the DC wires is used to measure the ionization losses, $dE/dX$, of charged
particles with $\sigma_{dE/dX}\approx$11-14\% accuracy for the minimun-ionizing particles.
 It is important for the analysis
to reconstruct a decay vertex position (the decay length of \Ks~ is
about 6 mm for the considered energies), and DC provides about 0.3 mm resolution
for the decay point radius.  

The barrel liquid-xenon (LXe) calorimeter with a 5.4~X$_0$ thickness has
fine electrode structure, providing a 1$\div$2 mm spatial resolution 
for photons~\cite{lxe}, and
shares the cryostat vacuum volume with the superconducting solenoid.     
The barrel CsI-crystal calorimeter is placed outside  the LXe calorimeter, 
and  increases the total thickness to   13.5~X$_0$.  The end cap BGO 
calorimeter with a thickness of 13.4~X$_0$ is placed inside the 
solenoid~\cite{cal}.
The luminosity is measured using events of Bhabha scattering 
at large angles with about 1\% accuracy~\cite{lum}. 

To understand the detector response to the processes under study we use Monte Carlo (MC) 
simulation of our detector based on the GEANT4~\cite{geant4} package, 
in which all simulated events pass the whole reconstruction and selection 
procedure. The MC simulation uses primary generator with matrix elements 
for the studied process, including soft photon radiation by the initial 
electron or positron~\cite{kur_fad, mcgpj}. For each energy point we
simulate 500 thousand signal events, and include experimental conditions
(efficiency, calibrations, beam position and energy etc.)  for the
detector elements. The neutral kaon mass in the simulation is set to
the PDG~\cite{pdg} value: m(\Kn) = 497.611~\mevcc.

%The data have been collected in the 2018 energy scan at 15 \Ecm~ 
%points around the $\phi$ resonance with 14.339 pb$^{-1}$ of the integrated luminosity.

%
\begin{figure}[tbh]
\begin{center}
%\vspace{-0.2cm}
\includegraphics[width=1.0\textwidth]{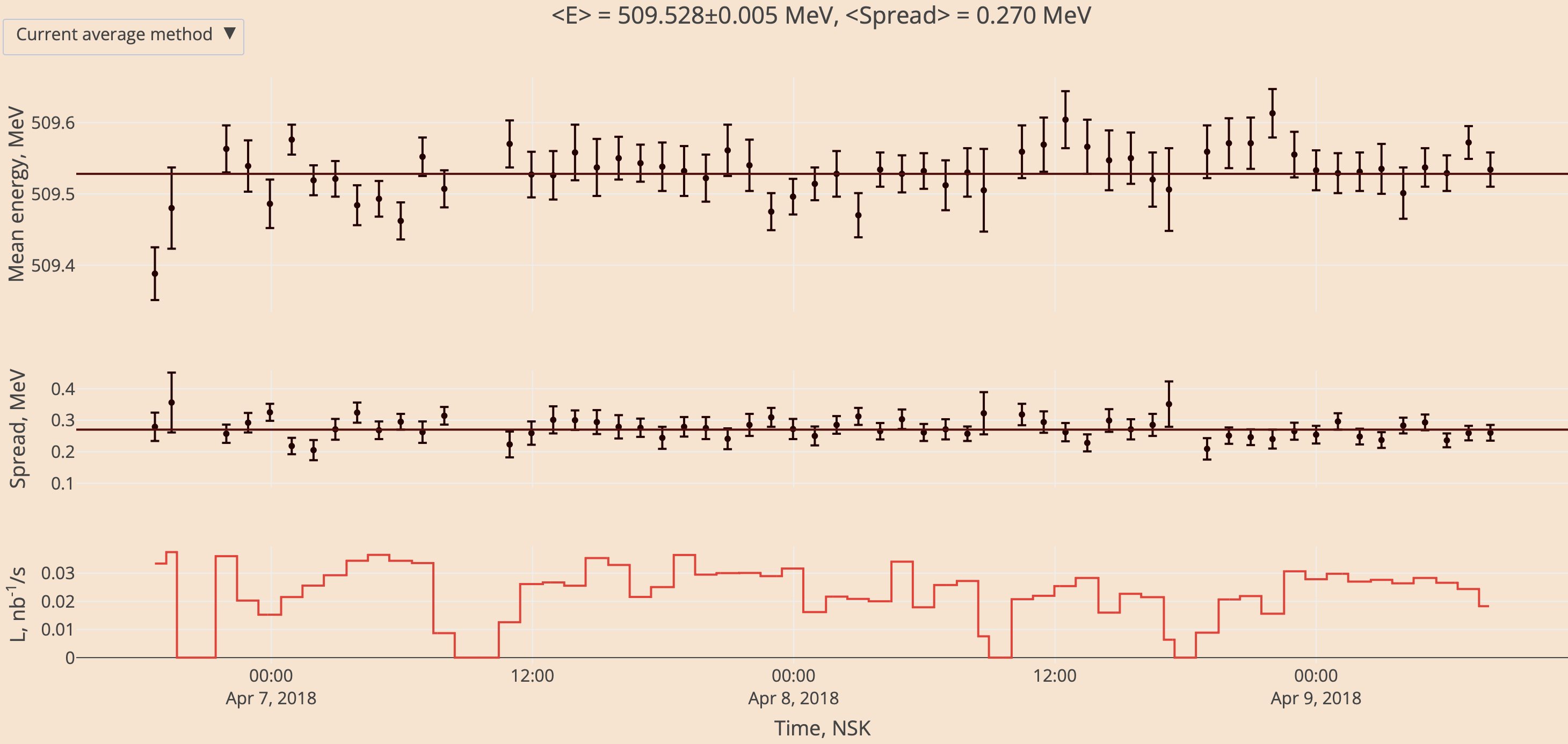}
%\vspace{-0.5cm}
\caption{
Data taking history for  \Eb=509.528 MeV:  the measurements of
the energy, the energy spread and the luminosity vs time are shown. Lines show average values.
}
\label{compt}
\end{center}
\end{figure}
\section{\boldmath Kaon energy}
\subsection{\boldmath Beam energy control}
\label{beamenergy}
\hspace*{\parindent}
The VEPP-2000 collider has one ring with the same orbit for the electrons
and positrons.
Because of the two-body $\phi$ decay
each neutral kaon has energy, close to the beam one, \Eb, with some corrections,
described  below. 
The beam energy is monitored using 
the Back-Scattering-Laser-Light system~\cite{laser}. During the data
taking for this experiment the beam energy measurements were performed every 20 minutes 
with the statistical uncertainty $\sigma\approx$30  keV.
The collider operators monitored the energy measurements
   and ensured that the machine operated within $\pm 2\sigma$ of the
   nominal energy   value.

 Figure~\ref{compt} shows a screenshot from the energy monitor
 presenting the  data-taking history for the \Eb=509.5 MeV.
 Each point corresponds to the beam energy and the energy
spread values, obtained from the back-scattered-photon energy spectrum fit as described
in ref.~\cite{laser}. The average values for the energy and
its  spread are also shown. If the measured energy deviates by more than
2$\sigma$ from the average, the corresponding data are not included into
analysis, like the first  point in figure~\ref{compt}:
less than one  percent of the events are removed. 
As a result, each experimental point has
the energy obtained by the averaging of the individual measurements with
much smaller statistical accuracy compared to the 30 keV systematic
uncertainty, declared in  ref.~\cite{laser}. The average also takes into
account a weight of each point, related to the luminosity integral,
shown at the bottom of figure~\ref{compt}. 

Another approach is presented in
the SND experiment for the $\phi(1020)$ meson parameters
study~\cite{sndphi}. It is  the most precise study of the $\phi$
resonance. The data in the the SND experiment were collected 
simultaneously  with CMD-3 using another collision section of the
collider, hence they have the same energies
and have luminosity integrals almost identical to those of CMD-3 for each energy
point. The average values for
the beam energy and the energy spread are obtained in SND study by the fitting of the 
combined back-scattering-photon energy spectra for each experimental
point. It allows them to obtain much better  statistical uncertainties for the energy spread. These
energies, \Eb, and the \Ecm~energy spreads, $\delta\Ecm$,  derived
from the SND paper~\cite{sndphi}, are listed in table~\ref{tabular}.

Also we should repeat a statement from the SND paper:

"The uncertainty of PDG~\cite{pdg} $\phi$-meson mass 1019.461$\pm$
0.016 MeV is significantly smaller than the systematic uncertainty of
the collider energy measurement (60 keV). Therefore, the PDG mass can
be used for calibration of the c.m. energy scale. To do this we introduce 
into the fit an additional parameter $\Delta\Ecm$ (common shift
of all energy points). Its fitted value
$\Delta\Ecm$ = 0.017$\pm$0.018 MeV''

The half of this value and uncertainty, $\Delta E_{SND} =
0.0085\pm0.0090$ MeV, is used to correct the beam energies in the first column of
table~\ref{tabular}. This correction value has been obtained with the SND energy set
therefore we use that for our analysis.

The energies obtained by averaging the
individual measurements (figure~\ref{compt}) do not differ significantly from those
listed in  table~\ref{tabular} , and are used for the systematic uncertainty study.

\subsection{\boldmath Kaon energy corrections}
\label{phienergyspread}
\hspace*{\parindent}
Current VEPP-2000 optics with the round beams at the interaction
region results in a relatively large collision energy spread, which accounts
for almost 10\% of the $\phi$-resonance width. At the sharp resonance slopes the
kaon pairs are produced with a larger cross section at one side of the
energy spread distribution and with a smaller cross section at the other
side. Therefore the average c.m. energy  of the kaons,
weighted with the resonance curve,
is shifted from the measured \Ecm~ energy.

Using the \Ecm~ energy spread from table~\ref{tabular} and the
$\phi$-meson parameters we convolute the resonance curve with the
Gaussian distribution of the beam energy around nominal \Ecm, and
calculate the weighted c.m. energy for the kaon pair at each
experimental point. The weighted c.m. energy is shifted up to $\pm$60 keV at the slopes,  and
the shift is zero on top of the resonance. 
We check also how large is the energy shift if we vary the
$\phi$-meson parameters or the energy spread value.

\begin{figure}[tbh]
\begin{center}
%\vspace{-0.5cm}
\includegraphics[width=1.0\textwidth]{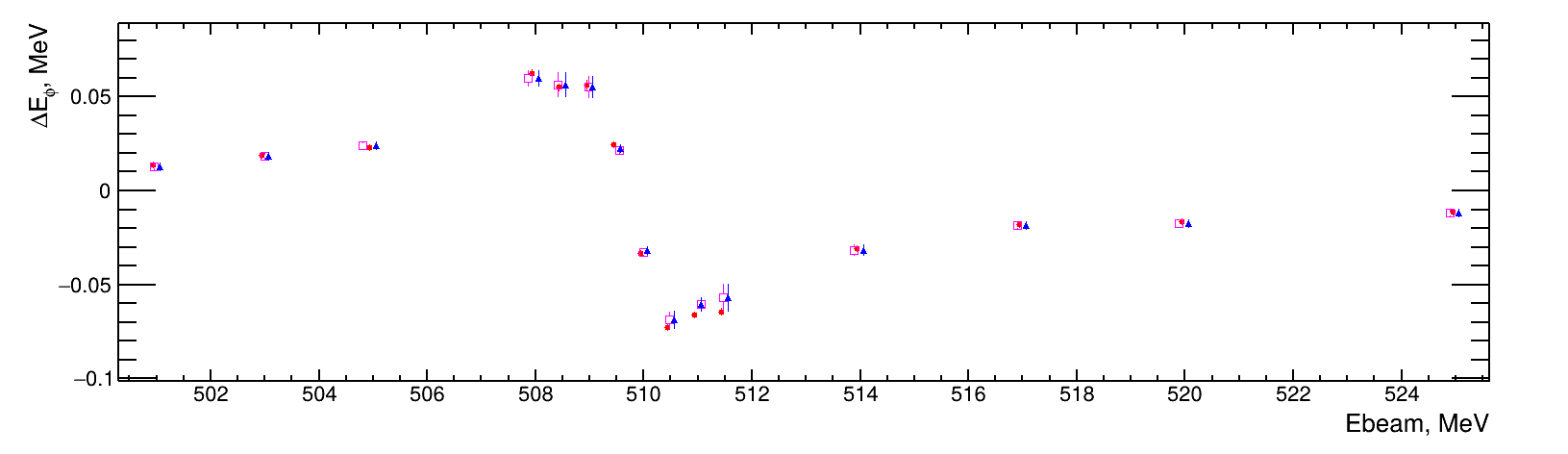}
\vspace{-0.5cm}
\caption{
The difference of the weighted c.m. energy of the kaons and the measured
\Ecm, $\Delta E_{\phi}$, for
different values of the  $\phi$-meson parameters and energy spreads:
triangles are for parameters from the PDG~\cite{pdg}, squares from the
SND measurement~\cite{sndphi}, and circles are for the averaging as in figure~\ref{compt}.
Point are separated by $\pm$0.06 MeV for visibility.)
}
\label{ekscorr}
\end{center}
\end{figure}

Figure~\ref{ekscorr} show  the difference, $\Delta
E_{\phi}$,  of the weighted energy from the measured \Ecm~
energy, obtained for the different conditions.   The error bars
show the uncertainty which comes from the accuracy of the energy spread measurements.
The triangle (blue) markers show the
obtained corrections if we use the $\phi$ width value from the
PDG~\cite{pdg}. The calculation with parameters
from the SND measurement~\cite{sndphi} are shown by the open squared (pink)
markers. The circular (red) markers are for the 
set of the energy spreads, obtained by averaging the individual measurements, as
shown in figure~\ref{compt}. The $\Delta E_{\phi}$ values for the
described conditions don't change by more
than 0.002--0.003 MeV, taken as an estimate of the systematic uncertainty.

The kaon energy is shifted by a half the presented difference,
$\Delta E_{\phi}$, and  the 
used corrections are listed in table~\ref{tabular}.

\section{\boldmath Signal events selection}
\hspace*{\parindent}
We select events of the $\epem\to\phi\to\Ks\Kl$ reaction, where the neutral kaon candidate
is reconstructed from the $\Ks\to\pipi$ decay to obtain
a clean event sample for the mass measurement.

At the first stage, our reconstruction procedure provides momentum,
angles, and dE/dX for each charged track in the DC.
We require exactly two oppositely charged
"good" tracks, each track has more than 10 hits in DC, its polar angle is in the
 range 1.0$\div$($\pi$-1) radians (tracks cross all
wires in DC), and the dE/dX values correspond to that of the charged pions.
We also require each track to be originated from the interaction region
along the beams within $\pm$10 cm from the center of the  interaction region.  

At the second stage, the pair of the  tracks is tested for the presence of a common vertex within
the DC resolution, and the invariant mass of these tracks, assuming them
to be pions, is required to be in the |m($\pipi$) -
m($K^{0}$)|<25~\mevcc~ range (about eight standard deviations from average). We
require the radius of the vertex position to be inside the vacuum beam
pipe of the collider.

\begin{figure}[tbh]
\begin{center}
%\vspace{-0.5cm}
\includegraphics[width=0.49\textwidth]{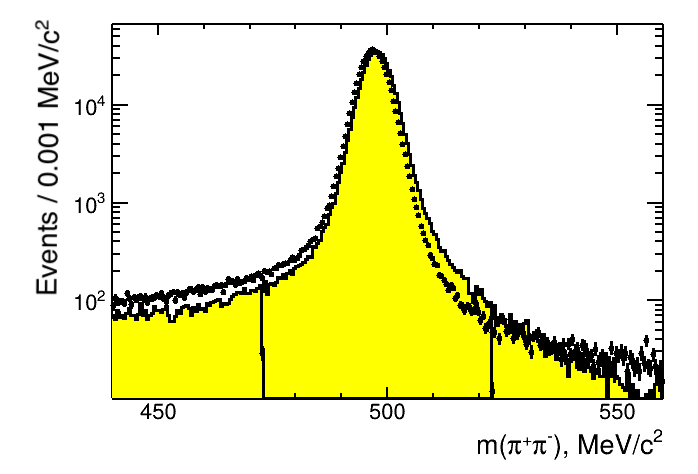}
\put(-50,100){\makebox(0,0)[lb]{\bf(a)}}
\includegraphics[width=0.49\textwidth]{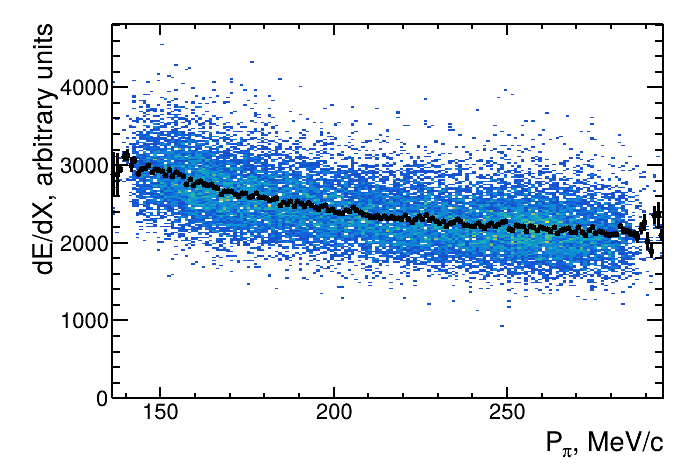}
\put(-50,100){\makebox(0,0)[lb]{\bf(b)}}
\vspace{-0.5cm}
\caption{
(a)  The invariant mass distribution for the two pions with the common
vertex for data (points) and  simulation (shaded histogram). Vertical lines show event selection for the mass measurement.
(b) The DC measured dE/dX vs pion momentum for the signal events. Points with
error bars show a profile  used for a momentum correction, described
in the text. 
}
\label{kinvmass}
\end{center}
\end{figure}

Figure~\ref{kinvmass}(a) shows the invariant mass for the pair of the  charged
pions with found common vertex for data (points) and simulation
(shaded histogram). Vertical lines show selections of events for
further analysis. In total we select 2039879 events corresponding to
the $\Ks\to\pipi$ decays with a practically negligible background. 
As shown in our previous paper on the $\epem\to\phi\to\Ks\Kl$ cross
section study~\cite{Kozyrev}, the contribution from other decays of the
$\phi$ resonance does not exceed a few per mil level, and do not provide a
peaking background  under the \Ks~ mass. The MC sample uses the world-average
kaon mass value~\cite{pdg} as input. The data and MC
distributions differ slightly in both the resolution, as discussed later, and the
peak position.

Figure~\ref{kinvmass}(b) presents a scatter
plot of the dE/dX values measured in DC vs the momenta of pions from the kaon decays.
The plot demonstrates  that the pions from the \Ks~
decay have momenta in the 140--290 MeV/c range were the energy loss in
the detector material is changing quite significantly.
 A shape, shown by the profile is used to estimate influence of this
effect to the result.

\section{\boldmath Edge angle approach using full two-body reconstruction}
\hspace*{\parindent}
As mentioned in the Introduction, the kaon mass can be calculated by
using the kinematical relation suggested in ref.~\cite{Zaitsev}. 
If the  pion 3-momenta, ${\bf p}_{+}, {\bf p}_{-}$, from the  $\Ks\to\pipi$ decay  are reconstructed,
the kaon mass value, ${\rm m}_{\Ks}$,  can be extracted from  the equation:
\begin{equation}
  \beta^{2}_{\Ks} = \frac{1}{\eta^{2}}(1+{\rm cos}\psi\sqrt{1-\eta^2})(1-\sqrt{1-\beta^2_{\pi}\eta^2}),
  \label{masseq}
\end{equation}
where $\eta = (1-Y^2)/(1+Y^2), Y=|{\bf p}_{+}|/|{\bf p}_{-}|$,~ 
  $\beta^{2}_{\Ks} = 1-({\rm m}_{\Ks}/{\rm E}_{\Ks})^2
  ,    ~~~\beta^2_{\pi}  = 1-(2{\rm m}_{\pi}/{\rm E}_{\Ks})^{2}$,
  ${\rm E}_{\Ks}$ is the kaon energy, and $\psi$ is the opening angle
  between the pions. The mass value in this equation depends on the kaon
  energy (assumed to be measured precisely), opening angle between pions,
  and the pion momenta ratio, $Y$; in the ratio it is reasonable to
  assume that a significant part of the systematic
  uncertainties related to the momentum measurement  cancels out.
For Y=1 eq.~\ref{masseq} transforms to  eq.~\ref{kmass0}.

\begin{figure}[tbh]
\begin{center}
\vspace{-0.2cm}
\includegraphics[width=0.34\textwidth]{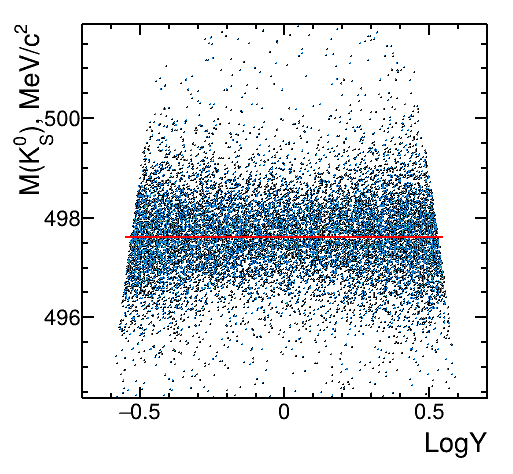}
\put(-50,100){\makebox(0,0)[lb]{\bf(a)}}
\includegraphics[width=0.34\textwidth]{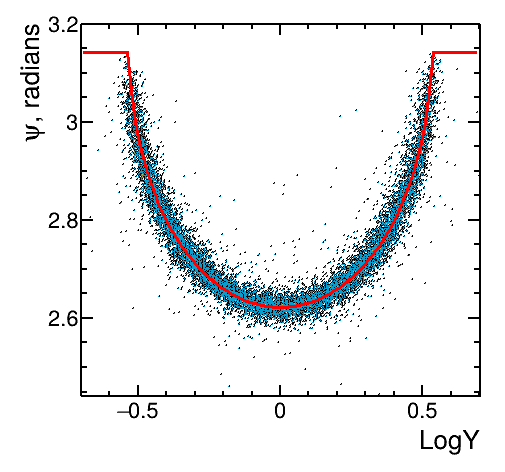}
\put(-50,100){\makebox(0,0)[lb]{\bf(b)}}
\includegraphics[width=0.34\textwidth]{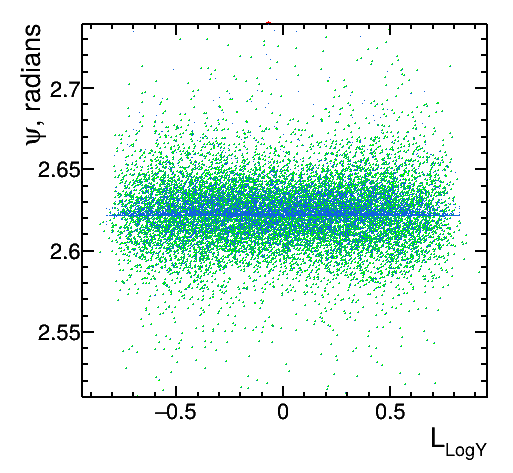}
\put(-50,100){\makebox(0,0)[lb]{\bf(c)}}
\vspace{-0.5cm}
\caption{
(a) The kaon mass for MC events with \Eb=509.5 MeV vs log(Y) calculated by eq.~\ref{masseq}. 
(b) The opening angle for MC pion pairs after reconstruction vs LogY  for \Eb=509.5 MeV with
the curve from eq.~\ref{masseq}.
(c) The modified opening angles vs curve length for the events in (b): see
explanation in the text. 
The narrow horizontal  structure corresponds to the events when the
"true'' generated angles are used.
}
\label{kmassY}
\end{center}
\end{figure}

Figure~\ref{kmassY}(a) shows the kaon mass vs log(Y), calculated for
the reconstructed events according to
eq.~\ref{masseq} for MC with \Eb=509.5 MeV. The log(Y) = log($p_{+}/p_{-}$), denoted as ${\text{LogY}}$,  is
a convenient parameter: LogY=0 if $p_{+}=p_{-}$, and has a positive
value when $p_{+}$ moves forward relative to the kaon direction
($p_{+}>p_{-} $) and  a negative value in the opposite case.   
The line shows the kaon mass value used in the simulation.
The experimental mass is obtained by  averaging of
the  profile points over a  LogY range: the procedure is used for the result described in
ref.~\cite{Zaitsev}. Unfortunately, the mass resolution is not uniform over LogY , and
a correction for the systematic
shift in each bin of the profile, reflecting the effect of the resolution on the  $\psi$ value, should
be applied. The resolution and distortion of the Y ratio, in which not all systematic uncertainties
are canceled out, are also essential.  Our study shows many systematic
effects which are limit the final precision.

The mass calculation using simple eq.~\ref{kmass0} is free from many
systematic effects, but needs precision measurement of the edge
angle.  Open angle between pions can be calculated using the same
kinematical relation as eq.~\ref{masseq}.  If the kaon energy and mass
values are used as input, 
the opening angle $\psi$ can be calculated vs LogY, and
the resulting curve is shown in figure~\ref{kmassY}(b).
The edge angle, $\psi_c^0$,   is the lowest point on the curve which
corresponds to LogY=0.

In the first collider experiments~\cite{barkov85, barkov87}, the
requirement that the relative difference of the pion momenta is less
than 10\% selects events  close to the edge angle.
  This requirement, which
corresponds to |LogY|$<\approx$0.1 in figure~\ref{kmassY}(b), 
highly reduces the statistics, but still saves events not at the edge angle.  

The experimental points, shown in figure~\ref{kmassY}(b), are distributed
around the calculated  "base line", reflecting the uncertainties in the $\psi$
and the LogY values after reconstruction.   
The idea is to use the "base line" behaviour to compensate the deviations of
the experimental points from the $\psi_c^0$ value for all measured
angles. 

We develop a procedure which minimizes the mean square deviation (using
weights according to the $\psi$ and LogY resolutions) of a given experimental point
on the two-dimentional distribution in figure~\ref{kmassY}(b) from the point on the
curve, and calculates a length, L$_{\rm  LogY}$, of the curve segment from LogY=0 to the
 determined optimal point (with sign) in the units of LogY.
A "modified opening angle" for each measurement is calculated as a sum of
the deviation value and the $\psi_c^0$ value. The mean square deviation
can be positive and negative depending on the side from the curve, and
correctly takes into account the resolutions in angle and LogY.
Figure~\ref{kmassY}(c) shows a scatter plot of the modified
opening angles vs L$_{\rm LogY}$. The measured points are relatively
uniformly distributed along the theoretical curve, and the
 length  L$_{\rm LogY}$ is a "natural'' parameter of the points along X
axis.

It is important that the procedure does not shift the positions of the
measured points close to the edge angle, where 
LogY=0, even if the input mass  in eq.~\ref{masseq}  deviates
 from the PDG value; however, it moves all the other points to be
distributed around  the edge angle value.

We check the procedure by applying it to the "true'' open angles of the pions from the primary
generator, and obtain the dark points in figure~\ref{kmassY}(c) which  form a
sharp narrow structure exactly at the $\psi_c^0$  edge angle value.
 Note that for the generator angles there are no
points below the $\psi_c^0$ value, but large angles are allowed, because 
the radiative effects decrease the kaon energy. 

\begin{figure}[tbh]
\begin{center}
\vspace{-0.2cm}
\includegraphics[width=0.495\textwidth]{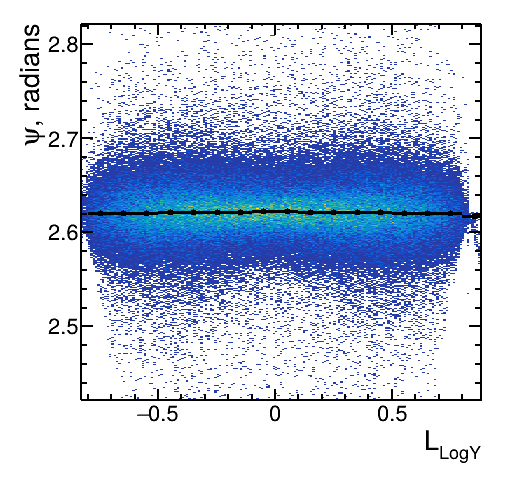}
\put(-30,170){\makebox(0,0)[lb]{\bf(a)}}
\includegraphics[width=0.5\textwidth]{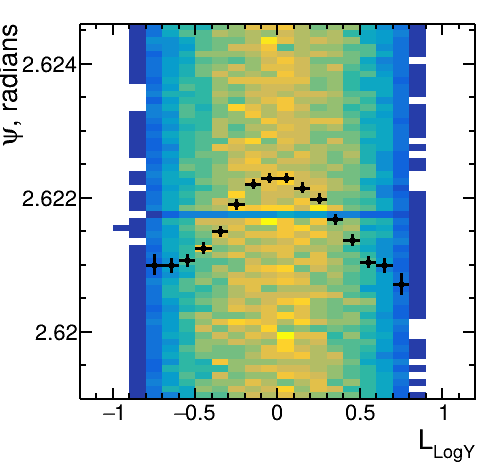}
\put(-30,170){\makebox(0,0)[lb]{\bf(b)}}
\vspace{-0.5cm}
\caption{
  (a) The modified  opening angles vs length distribution for the MC events  at  \Eb=509.5 MeV with the
  profile points before the corrections. 
(b) The expanded view of (a).
}
\label{dpsiprofMC}
\end{center}
\end{figure}
\begin{figure}[tbh]
\begin{center}
\vspace{-0.2cm}
\includegraphics[width=0.48\textwidth,height=0.465\textwidth]{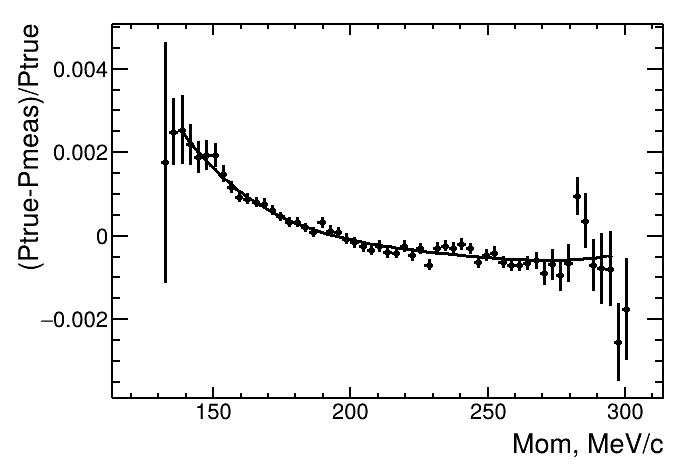}
\put(-30,150){\makebox(0,0)[lb]{\bf(a)}}
\includegraphics[width=0.5\textwidth]{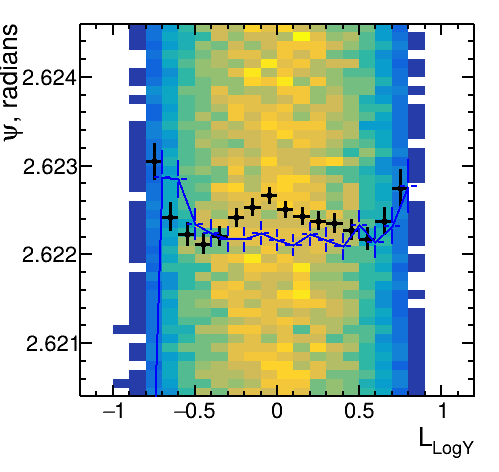}
\put(-30,150){\makebox(0,0)[lb]{\bf(b)}}
\vspace{-0.5cm}
\caption{
  (a)  The relative difference of the generated and the reconstructed momentum
  of the pions vs generated momentum  in the studied momentum range with a fit function, used for the correction.
  (b) The expanded view of the modified edge angle for the MC events with
  \Eb=509.5~MeV after correction. The results of the 
  "standard'' profile,  black points,  and the Gaussian profile, blue
  connected crosses,  are shown.
}
\label{dpsicorrMC}
\end{center}
\end{figure}

\section{\boldmath Edge angles in MC}
\subsection{\boldmath Edge angle extraction procedure and corrections}
\label{edgecorrMC}
\hspace*{\parindent}
After the described transformation, the modified angle distribution of
figure~\ref{kmassY}(c) is 
subjected to a profile procedure to get the average edge angle
value vs L$_{\rm LogY}$, as shown in figure~\ref{dpsiprofMC}(a). The  "standard'' profile, which
calculates the  average values of the points in a $\pm$0.15 radians window, is much more reliable
than that for the distribution in figure~\ref{kmassY}(a) for the kaon mass. But an
expanded view of this plot,  figure~\ref{dpsiprofMC}(b), shows a significant deviation of the
profile values from the input edge angle, corresponding to the
horizontal line. As shown below, 1 mrad in the egde angle value
corresponds to about 50 keV/c$^{2}$ in the kaon mass.

This deviation is caused by a rapid change of the ionization energy
loss of the pion vs its momentum (see figure~\ref{kinvmass}(b)) in the
detector material, mostly in the beam pipe,   resulting a non-linearity of the Y value.
A relative difference  of the
generated, P$_{\rm true}$, and the reconstructed, P$_{\rm meas}$, pion momentum vs the generated pion
momentum is shown in figure~\ref{dpsicorrMC}(a), demonstrating the
energy loss effect.
Note that a similar comparison of the generated and the reconstructed angles
of the pions does not show a significant difference.
We correct each reconstructed
pion momentum in the event according
to the second order polynomial fit function, shown in
figure~\ref{dpsicorrMC}(a).  The resulting
"standard'' profile is shown in figure~\ref{dpsicorrMC}(b) by the black
points: the deviations are smaller but still not uniform. 

As already pointed above, the result of the "standard'' profile very much depends on
the region of averaging, asymmetry in resolution, and on  the background
distribution tails. 
The profile result depends also on  the distribution tails from the
emission of the the soft photons by the initial particles.

To minimize these effects, we use an symmetric  "Gaussian 0.1--0.5
fit'' in the  profile.
The "Gaussian 0.1--0.5 fit''  is an iteration procedure which 
 is adjusting the fitting interval into the range where the Gaussian curve
 cover the distribution from the level 
 of 0.1 of the maximum value  below the 
 peak position to the level of 0.5 above.
 The adjusted interval corresponds to approximately
 -2.14$\sigma$  to 1.18$\sigma$ range  from the
 Gaussian maximum position.  The resulting Gaussian parameters  become
 stable enough after three-four iterations.  
 
We use the  "Gaussian 0.1--0.5 fit'' procedure in each 0.1 interval of the
L$_{\rm LogY}$ value for the points in figure~\ref{dpsiprofMC}(a).
The position values of the resulting Gaussian functions are shown by
the connected crosses in figure~\ref{dpsicorrMC}(b), and they 
exhibit a nearly uniform behaviour in a region around $\rm L_{\rm LogY}$=0.

\begin{figure}[tbh]
\begin{center}
\vspace{-0.1cm}
\includegraphics[width=1.02\textwidth,height=0.75\textwidth]{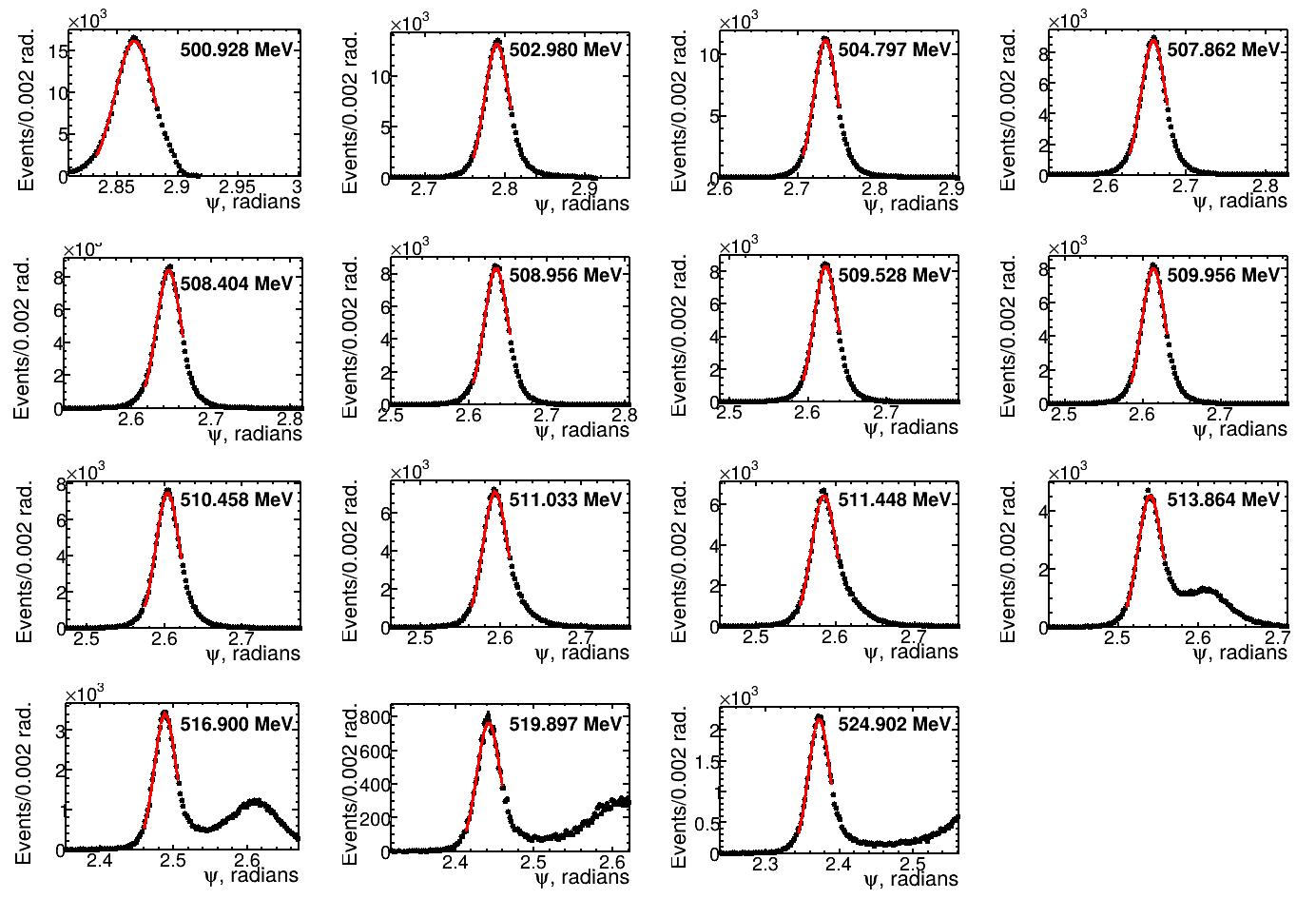}
\vspace{-0.5cm}
\caption{
The modified angle distributions for MC events with |L$_{\rm LogY} $|<0.2  at
  the different energies. Fit  functions, as described in the text,
  are superimposed on the MC data
}
\label{psifitMC}
\end{center}
\end{figure}
\subsection{\boldmath Fit to the edge angle in MC}
\hspace*{\parindent}
We perform the above corrections for all energy points of the simulated
events. A reasonably uniform behavior of the "Gaussian 0.1--0.5 fit'' profiles around 
L$_{\rm LogY}$=0 value is ensured.  We select events in the
|L$_{\rm LogY}$|<0.2 region, saving about 30\% of the signal, and make
projection plots to get distributions of the
events  concentrated at the edge angles.
We apply the "Gaussian 0.1--0.5 fit'' procedure to the distributions
for the energy points from \Eb=501 MeV to \Eb=525 MeV,
and figure~\ref{psifitMC} shows the result.
It is clearly seen how the soft photon emission extends distributions
to the larger angle values depending on the position on the $\phi$
resonance.
Fitting in the restricted angular region (from 0.1 to  0.5 of the
Gaussian maximum) reduces the influence of the radiation effects. 

\begin{figure}[tbh]
\begin{center}
  %\vspace{-0.2cm}
\includegraphics[width=0.334\textwidth]{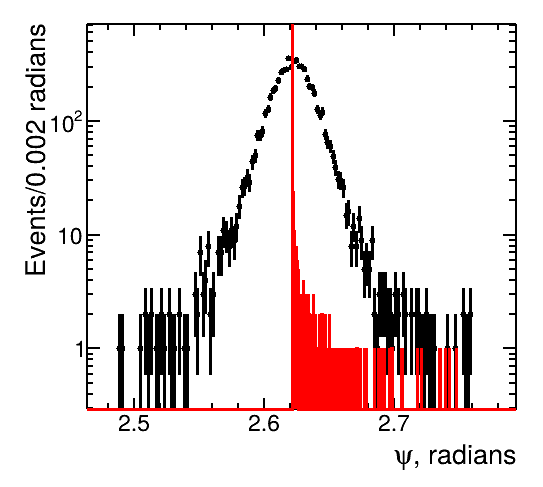}
\put(-50,100){\makebox(0,0)[lb]{\bf(a)}}
\includegraphics[width=0.323\textwidth]{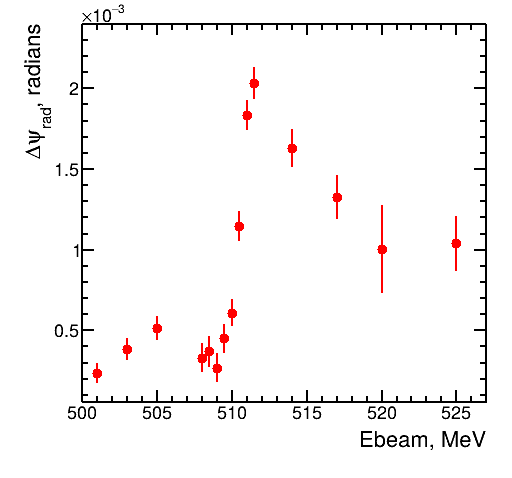}
\put(-50,100){\makebox(0,0)[lb]{\bf(b)}}
\includegraphics[width=0.33\textwidth]{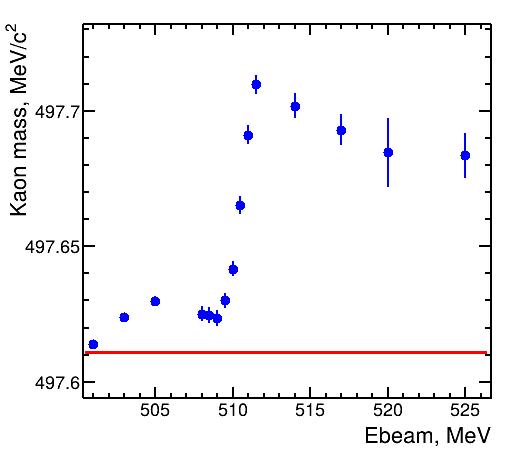}
\put(-50,100){\makebox(0,0)[lb]{\bf(c)}}
\vspace{-0.5cm}
\caption{
  (a) The modified angle distributions for the generated (red histogram) and
  reconstructed (black points)  events from simulation.
(b) The radiative correction for the edge angle after reconstruction.
(c) The calculated kaon mass from simulation using the fit values from
figure~\ref{psifitMC}. Line shows the mass value used in simulation.
}
\label{radangle}
\end{center}
\end{figure}

\subsection{\boldmath Radiative corrections to the edge angle}
\label{radcorr}
\hspace*{\parindent}
Figure~\ref{radangle}(a) shows a projection plot of the simulated
events in the  |L$_{\rm LogY}$|<0.2 region shown in figure~\ref{kmassY}(c) for the
modified reconstructed angles (black points)  and
for the modified generated angles (red histogram). The generated
angles have a sharp peak at the edge angle and a tail caused by the
radiation of the soft
photons. After  convolution with the detector resolutions, the average
of the reconstructed distribution, obtained from the "Gaussian 0.1--0.5 fit'', is
shifted from the generated edge angle: the shift is considered as a radiative
correction, $\Delta\psi_{rad}$, to the measured $\psi_c$ value. This
correction vs beam energy is shown in figure~\ref{radangle}(b) and
listed in table~\ref{tabular}.  To
demonstrate how large the correction is, we calculate the kaon
mass using eq.~\ref{kmass0} with the fitted values in
figure~\ref{psifitMC}, and show them in figure~\ref{radangle}(c). The line
shows the \Ks~ mass value used in the simulation.
The radiative correction values to the kaon mass are about
0.010$\div$0.020\,\mevcc~around the resonance peak 
and below, while they increase to about 0.100\,\mevcc~at higher energies. 

\begin{figure}[tbh]
\begin{center}
\vspace{-0.2cm}
\includegraphics[width=0.49\textwidth]{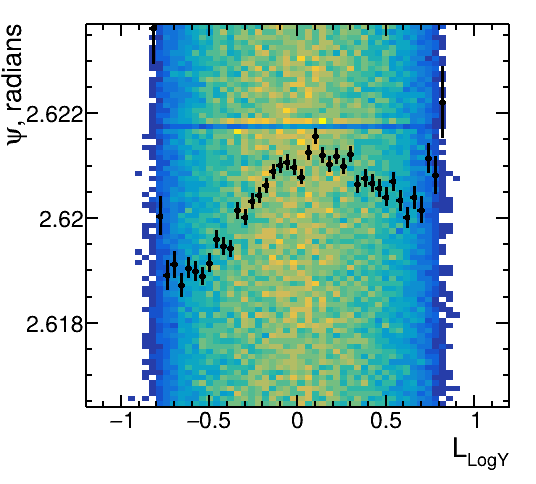}
\put(-30,150){\makebox(0,0)[lb]{\bf(a)}}
\includegraphics[width=0.5\textwidth]{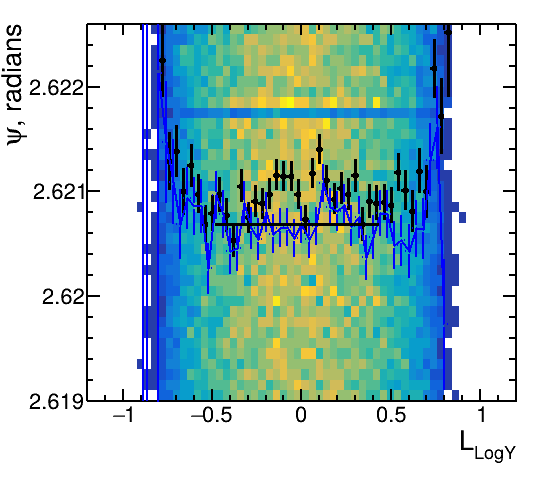}
\put(-30,150){\makebox(0,0)[lb]{\bf(b)}}
\vspace{-0.5cm}
\caption{
  (a) An expanded view for the  modified  angle vs length distribution
  for the experimental events at  \Eb=509.5 MeV with a "standard'' 
  profile points before corrections. 
(b) The events from  (a) after momentum corrections with the "standard''
  profile (points) and the Gaussian profile (connected crosses). Line
  shows a fit with a constant to the Gaussian profile points.
}
\label{dpsiprofData}
\end{center}
\end{figure}
\section{\boldmath Edge angles in data}

\subsection{\boldmath Edge angle corrections: pion momenta}
\label{edgecorrData}
\hspace*{\parindent}
 The expanded view of the scatter plot for the modified angles vs length for the
experimental  events at \Eb=509.5 MeV with  "standard" profile
points is shown in figure~\ref{dpsiprofData}(a) before any
corrections. The profile demonstrates a relatively large 
deviation from the uniform behaviour, similar to that from the
simulation, shown in figure~\ref{dpsiprofMC}(b).

The correction of each pion momentum according to the function shown in
 figure~\ref{dpsicorrMC}(a) almost completely removes the
observed large deviations.

The profile in figure~\ref{dpsiprofData}(a)  is also somehow
asymmetric relative to the center at L$_{\rm LogY}$=0. 
We investigate the influence of the other uncertainties in the momentum
measurement on the determination of the edge angle. 
The present  DC calibration procedure~\cite{dc}, based on the high energy
cosmic muons, has a limited precision,  
which  leads to a small difference in the
$p_{+}$ and $p_{-}$ values, and the edge angle is not at Y=1.
In addition to the momentum correction we introduce a shift to the
Y value at the level of 0.001, (one-per-mil for the difference in the momenta) which
removes the  asymmetry. 

 The events from figure~\ref{dpsiprofData}(a) after applying the corrections
 to the momenta and to Y value are shown in figure~\ref{dpsiprofData}(b) with
 the results of the "standard'' profile, and the result of our
 "Gaussian 0.1--0.5 fit''  profile. The profile points are well described by a constant function
in the -0.5$\div$0.5 region,  and
the resulting shift in the L$_{\rm LogY}$ value does not introduce more than 0.00002
radians uncertainty (about 1 keV/c$^2$ in mass). 
 
 It should be stated again - any manipulations with Y value
does not change the position of the modified angles around L$_{\rm
  LogY}$ = 0. Our corrections require that the  values  of the
"Gaussian 0.1--0.5 fit" profile in all other intervals do not deviate
from the edge value beyond the statistical uncertainty.

\begin{figure}[tbh]
\begin{center}
%\vspace{-0.2cm}
  \includegraphics[width=1.0\textwidth]{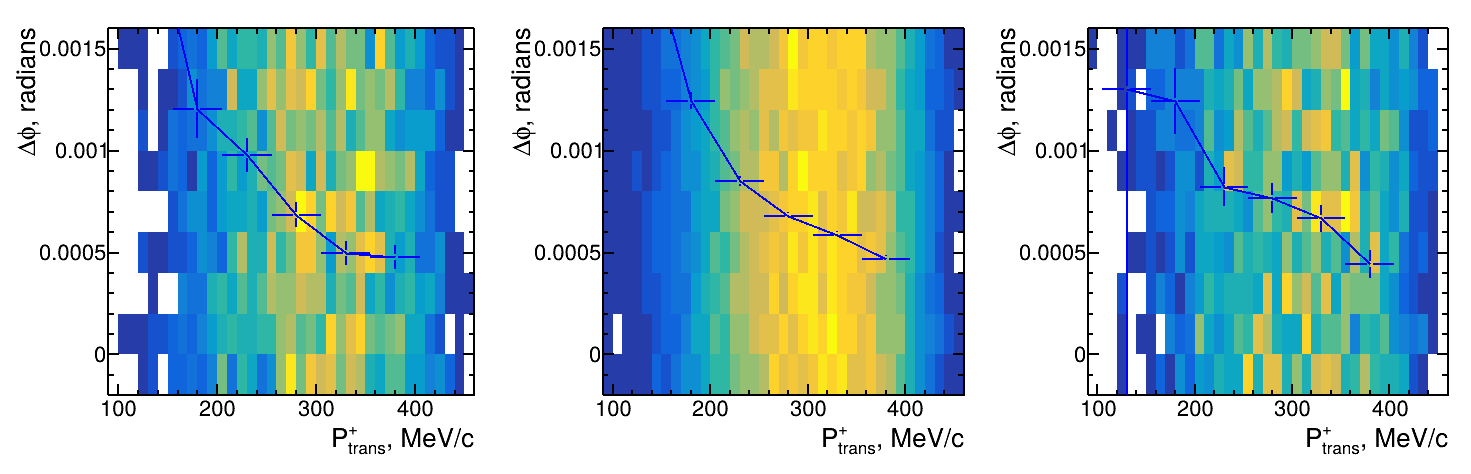}
  \includegraphics[width=1.0\textwidth]{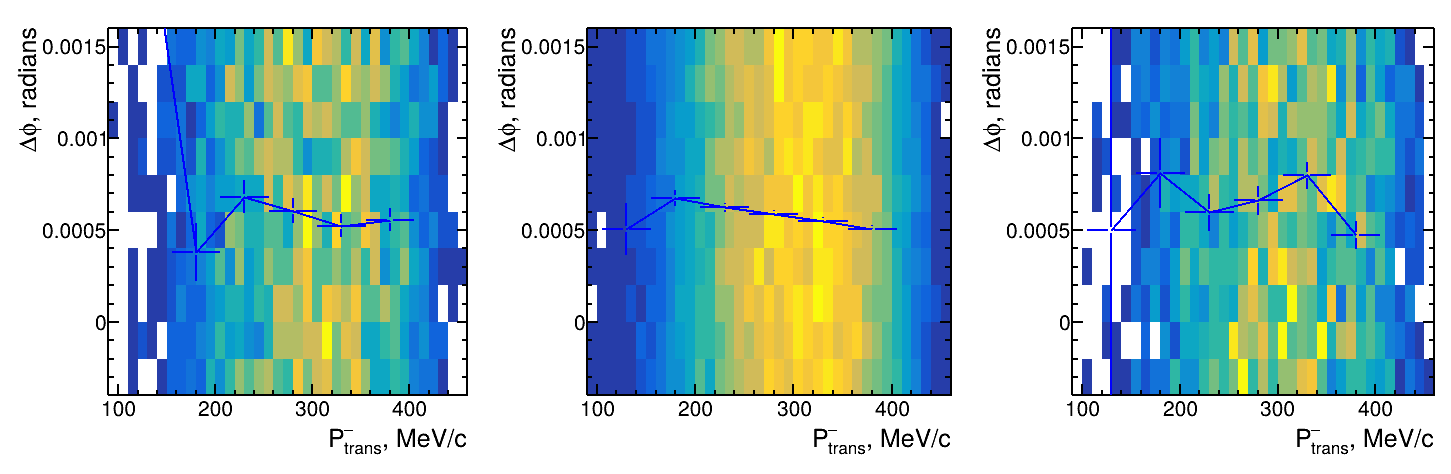}
\vspace{-0.5cm}
\caption{
  An  example of the profiles for the azimuth angle deviations for the positive (upper plots)
  and negative (lower plots) pions vs the  transverse
  momentum for the 508, 509.5 and 511.5 MeV beam energies.
}
\label{phicorr3pi}
\end{center}
\end{figure}
\subsection{\boldmath Edge angle corrections: pion angles}
\label{edgecorrDataAngles}
\hspace*{\parindent}
As noted in Sec.~\ref{edgecorrMC} the reconstructed angles of the pions
in simulation do not
deviate (on average) from the initial generator values. 

To investigate  possible deviations in data, we develop a procedure with 
test events from the $\epem\to\phi\to\pipi\piz$  reaction.
The pions from the $\phi\to\pipi\piz$ decay originate from the
collision region and have a
momentum range similar to that from the \Ks~ decay.  

The position of the collision point in a transverse plane is known with a few
microns accuracy, and the colliding beams have a transverse size of
about 10 microns: ten times better than the DC resolution.
For each track our reconstruction procedure fits the hits in the DC to
obtain  the production angles
 and momentum. In addition, we perform a fit which uses the
interaction point  as an additional hit on the track:  weights of
the  DC hits, starting at the radius of 4 cm, are significantly reduced, therefore
the reconstructed angle is closer to the true one.   The
difference in the azimuthal angles for these two fits provides an estimate
of the systematic angle shift.

A relatively clean
sample of the  $\phi\to\pipi\piz$ events  is selected by  requiring
energy-momentum conservation for the detected two oppositely-charged
pion tracks,  and two 
photons in the calorimeter with an invariant mass close to that of
the \piz.
We make a scatter plot of the difference between the two azimuthal
angles vs the transverse momentum,  separately for the  positive and the
negative pions. An expanded view of these scatter plots with
the Gaussian profiles for a few energy points is
shown in figure~\ref{phicorr3pi}. The azimuth angular differences, $\Delta\phi$, for the positive
and the negative pions have  different behaviours with the transverse
momentum, and the difference is not completely canceled when we
calculate the opening angle for the two tracks.

The \Ks~decay length is about 6 mm, and the pions from the decay are
somehow shifted from the collision point. The shifts shown in figure~\ref{phicorr3pi} are relatively
small, therefore  we assume similar angular differences $\Delta\phi$
vs momentum for the decay pions, and use them for corrections at each
energy point.

\begin{figure}[tbh]
\begin{center}
\vspace{-0.2cm}
\includegraphics[width=0.52\textwidth]{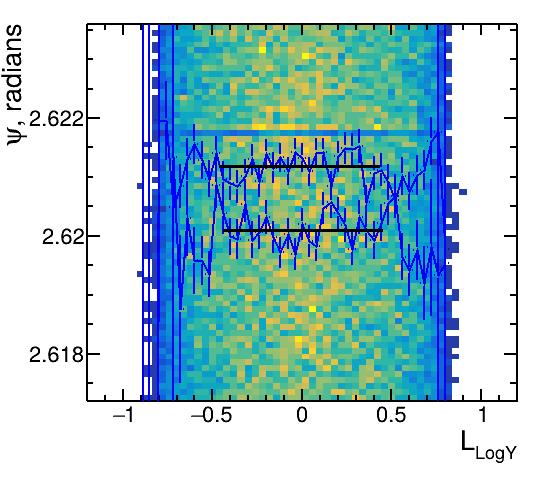}
\put(-30,170){\makebox(0,0)[lb]{\bf(a)}}
\includegraphics[width=0.5\textwidth]{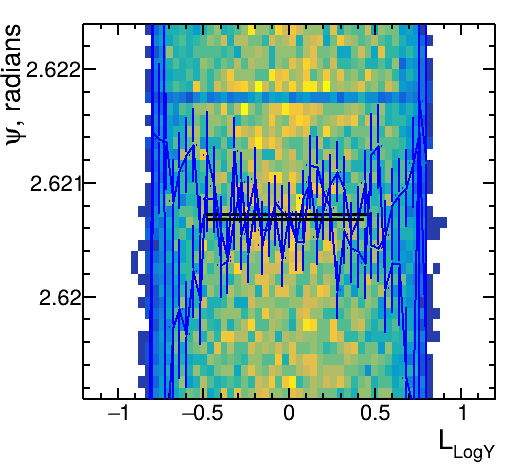}
\put(-30,170){\makebox(0,0)[lb]{\bf(b)}}
\vspace{-0.5cm}
\caption{
 The expanded view of the  angle vs length distribution for the experimenta events at
  \Eb=509.5 MeV with a Gaussian profiles for the "fish'' and the "bird'' type events
 before (a) and after (b) corrections with the fits explained in the text. 
}
\label{dpsiprofDataAngles}
\end{center}
\end{figure}

To study the influence of these effects on the edge angle determination, we divide
our data into two samples. 
One sample includes events for which the magnetic field curves pions inside the
opening angle between them, "fish" type events, and the second sample
includes pions moving
outside of the decay opening angle, "bird" type
events. Figure~\ref{dpsiprofDataAngles}(a) shows the 
scatter plot for the same events as in 
figure~\ref{dpsiprofData}(b) with the "Gaussian 0.1--0.5 fit'' profiles made separately for
the "fish" and the "bird" type events. The fits with constant yield 
2.62119$\pm$0.00006 and 2.62008$\pm$0.00005, respectively. The
difference of about 1 mrad is quite significant.

We correct the azimuth angle of each positive and negative pion track using the 
profiles shown in figure~\ref{phicorr3pi} for each  energy
point, and calculate the modified opening angles.
The results
for the two event types   from figure \ref{dpsiprofDataAngles}(a)  are shown in
figure~\ref{dpsiprofDataAngles}(b). 
The constant fits yield 2.62069$\pm$0.00006 radians and
2.62073$\pm$0.00005 radians, for the "fish" and "bird" type of the 
events, respectively. Note that the fit for  the  sum of the two samples after
corrections does not change by more than 0.00002 radians (about 1--2
keV/c$^{2}$ in the kaon mass), which is an estimate of the systematic
uncertainty of the procedure.  

\begin{figure}[tbh]
\begin{center}
\vspace{-0.1cm}
\includegraphics[width=1.02\textwidth,height=0.75\textwidth]{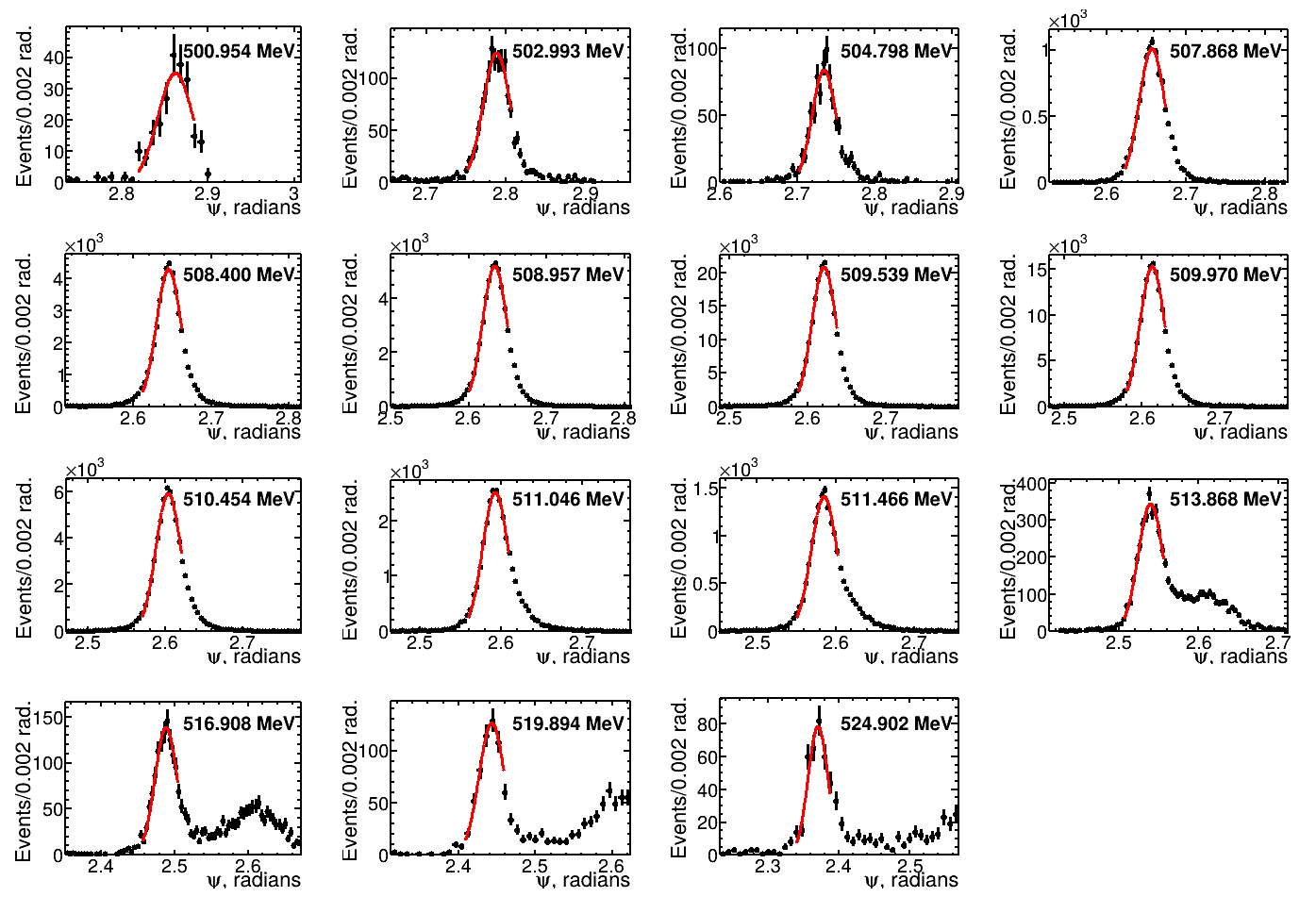}
\vspace{-0.5cm}
\caption{
  The edge angle distributions for events with |L$_{\rm LogY}$|<0.2 in data at
  different energies and the fit
  functions described in the text.
}
\label{psifitData}
\end{center}
\end{figure}
\begin{figure}[tbh]
\begin{center}
\vspace{-0.2cm}
\includegraphics[width=0.495\textwidth]{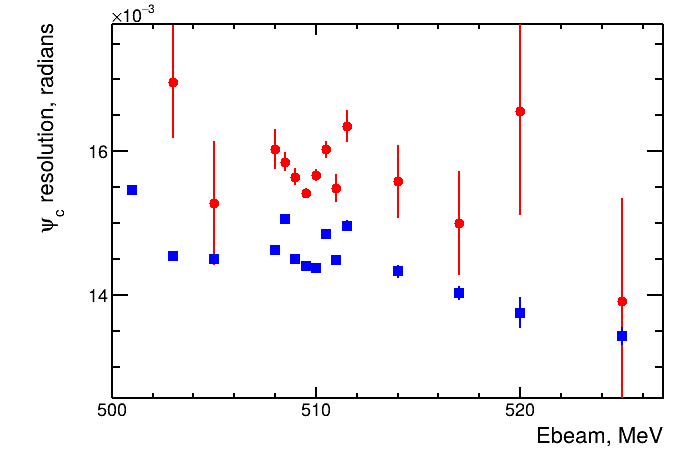}
\put(-50,100){\makebox(0,0)[lb]{\bf(a)}}
\includegraphics[width=0.50\textwidth]{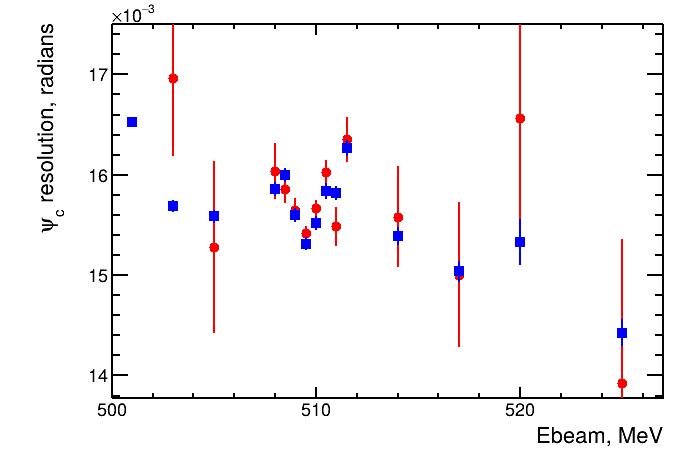}
\put(-50,100){\makebox(0,0)[lb]{\bf(b)}}
\vspace{-0.3cm}
\caption{
The  edge angle resolution for simulation (squares) and data (circles) obtained from
  the fits shown in figure~\ref{psifitMC} and figure~\ref{psifitData} before (a) and
  after (b) an additional 4 mrad Gaussian convolution applied to the MC distributions
}
\label{psires}
\end{center}
\end{figure}
\subsection{\boldmath Fit to the edge angles and resolution}
\hspace*{\parindent}
Similar to the simulation, in data we select events in the |L$_{\rm LogY}$|<0.2 region and
make a projection plot to obtain the distributions for the angles at
each energy point,
shown in figure~\ref{psifitData}. We perform the "Gaussian 0.1--0.5
fit'', and get the edge angles and 
resolution values for each energy point. The obtained edge angles are
listed in table~\ref{tabular}. In total, 609583 events are used for
the edge angle determination. 

Figure~\ref{psires}(a) shows the obtained angular resolutions for data
and simulation vs energy. The edge angular resolution is about 16 mrad
for data and slightly better in the simulation. The angle
corrections for the  "fish-bird'' type of events improve the data
resolutions by about a half of milliradian.

As discussed in
Sec.~\ref{radcorr}, the radiative correction to the edge angle depends on the
detector resolution, and we introduce a 4 mrad Gaussian convolution to
the simulated modified angles after reconstruction. After these convolutions the angular resolutions
in data and simulation are close at the level of statistical
uncertainties, and are shown in figure~\ref{psires}(b). The remaining
uncertainties in the radiative corrections are discussed below. The radiative
corrections to the $\psi_c$, shown in figure~\ref{radangle}(b) and listed
in table~\ref{tabular}, include the additional smearing.
\section{\boldmath Kaon mass calculation}
\label{kmasscalc}
\hspace*{\parindent}
The neutral kaon mass is calculated according to eq.~\ref{kmass0} with
corrections to the edge angle and the energy as:
\begin{equation}
m(\Ks) = \sqrt{(\Eb + \Delta E_{\phi} +
  \Delta E_{SND})^2sin^2((\psi_c - \Delta\psi_{rad})/2)+4m^2_{\pi}cos^2((\psi_c -
    \Delta\psi_{rad})/2)}. 
\label{kmass1}
\end{equation}
\begin{figure}[tbh]
\begin{center}
\vspace{-0.2cm}
\includegraphics[width=0.495\textwidth]{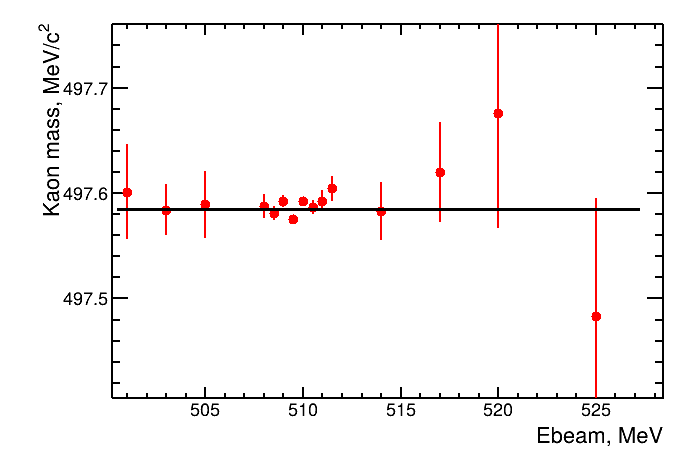}
\put(-40,110){\makebox(0,0)[lb]{\bf(a)}}
\includegraphics[width=0.5\textwidth]{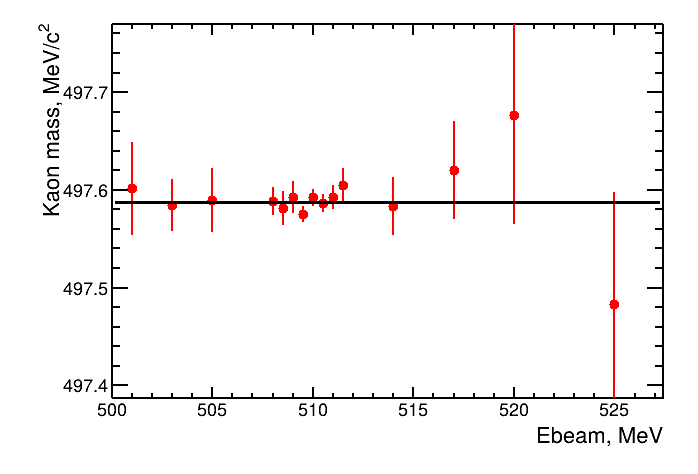}
\put(-40,110){\makebox(0,0)[lb]{\bf(b)}}
\vspace{-0.3cm}
\caption{
  (a) The  \Ks ~mass calculated for each energy point with only
  event statistic uncertainties included. The line shows the fit with constant value. 
  (b) The \Ks ~mass calculated for each energy point with all
  statistical uncertainties included. The line shows the fit with constant value.
}
\label{ksmass}
\end{center}
\end{figure}

Figure~\ref{ksmass} shows the kaon masses calculated for each energy
point with all corrections applied.    When only event statistical
uncertainties are included to the error
bars in figure~\ref{ksmass}(a), the fit with the constant yields the value
m(\Ks)=497.585$\pm$0.002 \mevcc~ with $\chi^2/n.d.f. = 15.8/14$.
The uncertainties in the energy measurements from table~\ref{tabular} are
uncorrelated with the event statistic and are also statistical in
nature. The fit with the constant value to the masses with all 
the statistical uncertainties included, shown in figure~\ref{ksmass}(b),
yields
$$
m(\Ks)=497.587 \pm 0.004 ~\mevcc~ with  ~\chi^2/n.d.f. = 5.8/14. 
$$
In the  result the statistical precision is  dominated  by  the uncertainties in
the energy measurements. 

The \Ks~mass values for each energy point are listed in
table~\ref{tabular}, and they have  an  uniform behaviour after
the relatively large corrections to the energies and angles. 

\section{\boldmath Systematic uncertainties}
\label{systematic}
\subsection{\boldmath Energy calibration}
\label{systenergy}
\hspace*{\parindent}
The uncertainty in the calibration constant $\Delta
E_{SND}=0.0085\pm0.009$ MeV, derived from the SND measurement~\cite{sndphi} of the
$\phi$-meson mass relative to the PDG world average value~\cite{pdg},   is the
major systematic uncertainty in our  kaon mass measurement, as shown
in table~\ref{tabsyst} (line 1). A work on the better
calibration of the laser system at the VEPP-2000 collider using the
resonance depolarization method is in progress, and could further
improve the result. 

The different approach to the average-over-point energy calculations,
discussed in Sec.~\ref{beamenergy}, gives  
a different result because of a possible small machine energy drift during the
data taking. For example, the average beam energy in figure~\ref{compt} is
\Eb=509.528$\pm$0.005 MeV
compared to the value  \Eb=509.539$\pm$0.008 MeV  from
table~\ref{tabular}, which is the largest observed difference.  
With another set of the energies
we obtain m(\Ks)=497.583$\pm$0.003\,\mevcc~ with $\chi^2/n.d.f. = 7.7/14$. 
The 0.004 \mevcc~ difference may be compensated in the $\Delta E_{SND}$ value if this energy set is
used in the $\phi$ mass measurement, but conservatively we 
take it as an estimate of the systematic uncertainty. It is  listed in
table~\ref{tabsyst} (line 2).

\subsection{\boldmath Energy shifts at $\phi$ resonance}
\label{phisift}
\hspace*{\parindent}
As discussed in Sec.~\ref{phienergyspread}, the beam energy spread
shifts the effective kaon energies because of the fast cross section
changing at the $\phi$ resonance. The corrections from -0.034 MeV to
+0.030 MeV are introduced, listed in table~\ref{tabular}. 
The uncertainties in the $\phi$-meson
parameters and in the energy spread give about 0.002\,MeV
variation in these numbers, which  is added as the systematic uncertainty for
the kaon mass and listed in  table~\ref{tabsyst} (line 3).

\subsection{\boldmath Edge angles vs LogY corrections}
\label{systegde}
\hspace*{\parindent}
We correct the irregularity in the modified edge angle distributions
vs L$_{\rm LogY}$, figures~\ref{dpsiprofMC},\ref{dpsiprofData}, 
using  physical effects like changing momentum behaviour because of
the energy loss in the detector material or uncertainty in the momentum
calibration (see Sec.~\ref{edgecorrMC},\ref{edgecorrData}). We also
try to apply  corrections directly to the Y value with a simple
function to make the resulting "Gaussian 0.1--0.5 fit''  profile
uniform around L$_{\rm LogY}$=0. The result in the kaon mass  changes
by less than 0.001\,\mevcc, which is added as a systematic uncertainty.

We also perform all calculations with
the selections |L$_{\rm LogY}$|<0.1 and |L$_{\rm LogY}$|<0.3: number  of
events are changed by factor of two.  The kaon masses
m(\Ks)=497.587$\pm$0.0045\,\mevcc\\ and
m(\Ks)=497.588$\pm$0.0039\,\mevcc~  have been obtained,
respectively. The 0.001\,\mevcc\\
difference is also added as a systematic uncertainty to the kaon mass.

The smallness of these uncertainties, listed in table~\ref{tabsyst}
(lines 4 and 5),  reflects the fact that the
modified edge angle is almost independent of Y in the vicinity of L$_{\rm LogY}$=0.

\begin{figure}[tbh]
\begin{center}
\vspace{-0.2cm}
\includegraphics[width=0.35\textwidth]{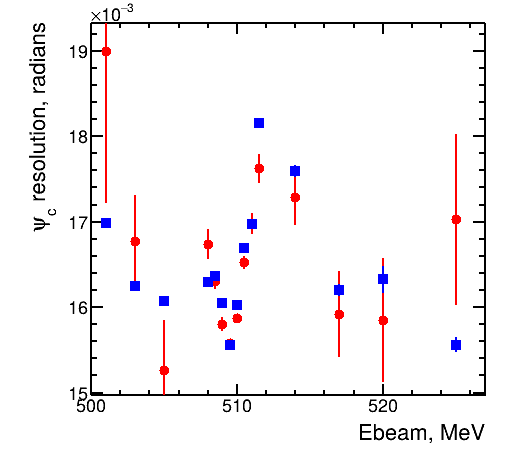}
\put(-50,100){\makebox(0,0)[lb]{\bf(a)}}
\includegraphics[width=0.34\textwidth]{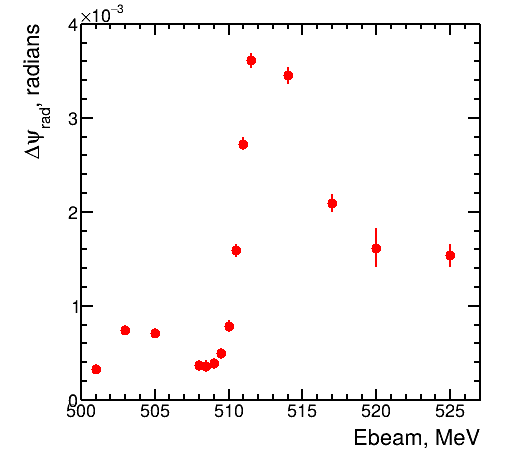}
\put(-50,100){\makebox(0,0)[lb]{\bf(b)}}
\includegraphics[width=0.33\textwidth]{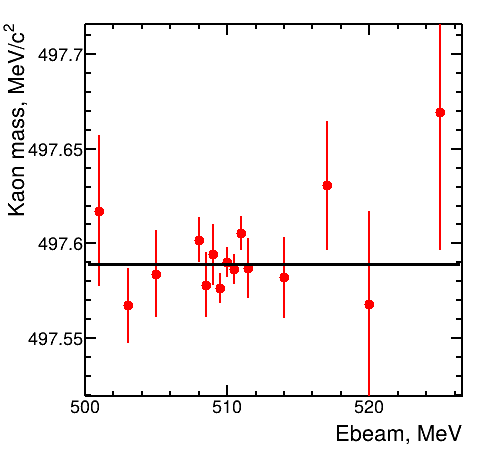}
\put(-50,100){\makebox(0,0)[lb]{\bf(c)}}
\vspace{-0.5cm}
\caption{
  (a) The edge angle resolution for the 0.1$\div$0.1 fits for data(circles) and MC(squares).
(b) The radiative correction to the edge angle for the 0.1$\div$0.1 fits.
(c) The kaon mass for the 0.1$\div$0.1 fits of the edge angles. 
}
\label{radcorr1}
\end{center}
\end{figure}

\subsection{\boldmath Radiative corrections}
\label{systrad}
\hspace*{\parindent}
As discussed in Sec.~\ref{radcorr} the radiative correction to the edge angle
depends on the detector resolution and varies from 0.09 to 2.07 mrad,
as shown in table~\ref{tabular}. It is the second largest
correction corresponding to  0.015$\div$0.100\,\mevcc~ in the kaon
mass.   We add the 4 mrad  Gaussian spread to the MC angular
distribution, which  increases angular resolution by about 1 
mrad and makes it close to that from data within a statistical
uncertainty of 0.2 mrad: see figure~\ref{psires}(b).
The 1 mrad change in the resolution corresponds to about
0.005\,\mevcc~  in the mass value at the resonance maximum, and  a remaining systematic
uncertainty is estimated as 0.001\,\mevcc ~(line 6 of table~\ref{tabsyst}).

The simulated events are produced with the  fixed  \Ecm~energies 
without any spread. The primary generator ~\cite{mcgpj} does not allow
us to introduce
the energy spread. As discussed above, the weighted energy of the
kaons is shifted by $\pm$60 keV, as shown in figure~\ref{ekscorr},
therefore the radiative effects are calculated to shifted energy.
This effect is  small on top and below $\phi$-resonance maximum, but
above the peak the energy shift changes the radiative correction by
not more than 0.002\,\mevcc, which is taken as systematic uncertainty
estimate (line 7 of table~\ref{tabsyst}).

Much stronger test is performed by changing the fit ranges  in figures~\ref{psifitMC},\ref{psifitData}
from the 0.1$\div$0.5 to 0.1$\div$0.1 levels from the Gaussian curve
maximum: we apply the  "Gaussian 0.1--0.1 fit'' procedure. Figure~\ref{radcorr1}(a) shows that the  angular
resolution is not monotonic with energy anymore (radiative tails are
involved), and the radiative corrections to the edge angles are increased by a
factor of two.  Figure~\ref{radcorr1}(b) shows this in comparison with
figure~\ref{radangle}(b), and the fit to the resulting mass 
 in figure~\ref{radcorr1}(c)  yields  m(\Ks)=497.590$\pm$0.004 \mevcc~ with $\chi^2/n.d.f. =
12.4/14$. The 0.003 \mevcc~ shift is conservatively  taken as an estimate of the
radiative correction systematic uncertainty and listed in
table~\ref{tabsyst} (line 8).
\begin{figure}[tbh]
\begin{center}
\vspace{-0.2cm}
\includegraphics[width=0.495\textwidth]{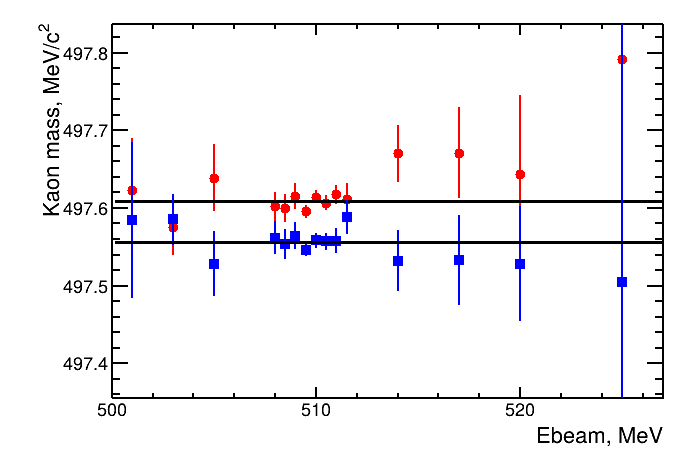}
\put(-40,110){\makebox(0,0)[lb]{\bf(a)}}
\includegraphics[width=0.5\textwidth]{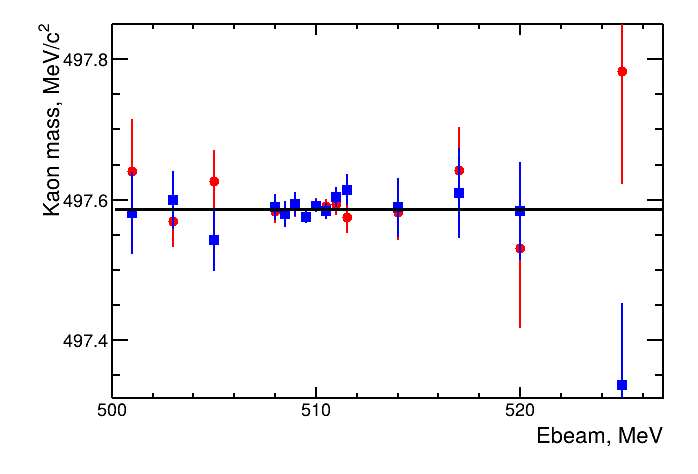}
\put(-40,110){\makebox(0,0)[lb]{\bf(b)}}
\vspace{-0.3cm}
\caption{
  (a) The  \Ks ~mass calculated for each energy point
  for the "bird'' (circles) and "fish'' (squares) type events. Lines show
  fits to the mass values.
  (b) The \Ks ~mass calculated for each energy point ater angle
  corrections for the "bird'' (circles) and "fish'' (squares) type
  events. Lines show the fit with a constant value.
}
\label{ksmassbirdfish}
\end{center}
\end{figure}
\subsection{\boldmath Angle corrections}
\label{systburdfish}
\hspace*{\parindent}
Figure~\ref{ksmassbirdfish}(a) shows the calculated \Ks~mass before
angle corrections for the "fish'' and the "bird'' type events. The fit
with constant value yields 
  m(\Ks)=497.609$\pm$0.0041 \mevcc~ with $\chi^2/n.d.f.=10.1/14$
and
 m(\Ks)=497.557$\pm$0.0045 \mevcc~ with $\chi^2/n.d.f.=5.4/14$,
 respectively. The mixture of the two type of events gives
m(\Ks)=497.585$\pm$0.0039 \mevcc~ with $\chi^2/n.d.f.=13.2/14$.

After applying the  corrections to the pion angles, the
fits yield \\  m(\Ks)=497.586$\pm$0.0044\,\mevcc~ with
$\chi^2/n.d.f.=7.1/14$ and \\
m(\Ks)=497.587$\pm$0.0045\,\mevcc~ with $\chi^2/n.d.f.=11.1/14$ \\for
the "fish'' and "bird'' type events, respectively. 

The small difference of 0.001\,\mevcc, listed in table~\ref{tabsyst}
 line 9), indicates a strong
cancellation of angular uncertainties when averaging over the ``fish''
and ``bird'' event  types.

We perform a much stronger test of the result. We require all charged
tracks to be in the $\pm$0.2 radians region perpendicular to the
beams axis, reducing influence of the events with a poor polar angle measurement.
This requirement reduces number of signal events by a factor
of two, and the resulting fit yields
m(\Ks)=497.583$\pm$0.0054\,\mevcc~ with $\chi^2/n.d.f.=18/14$. We
conservatively take the 0.004\,\mevcc~difference with final result as
an additional systematic uncertainty in the angle measurements, listed in
table~\ref{tabsyst} (line 10).

\begin{figure}[tbh]
\begin{center}
\vspace{-0.2cm}
\includegraphics[width=1.0\textwidth]{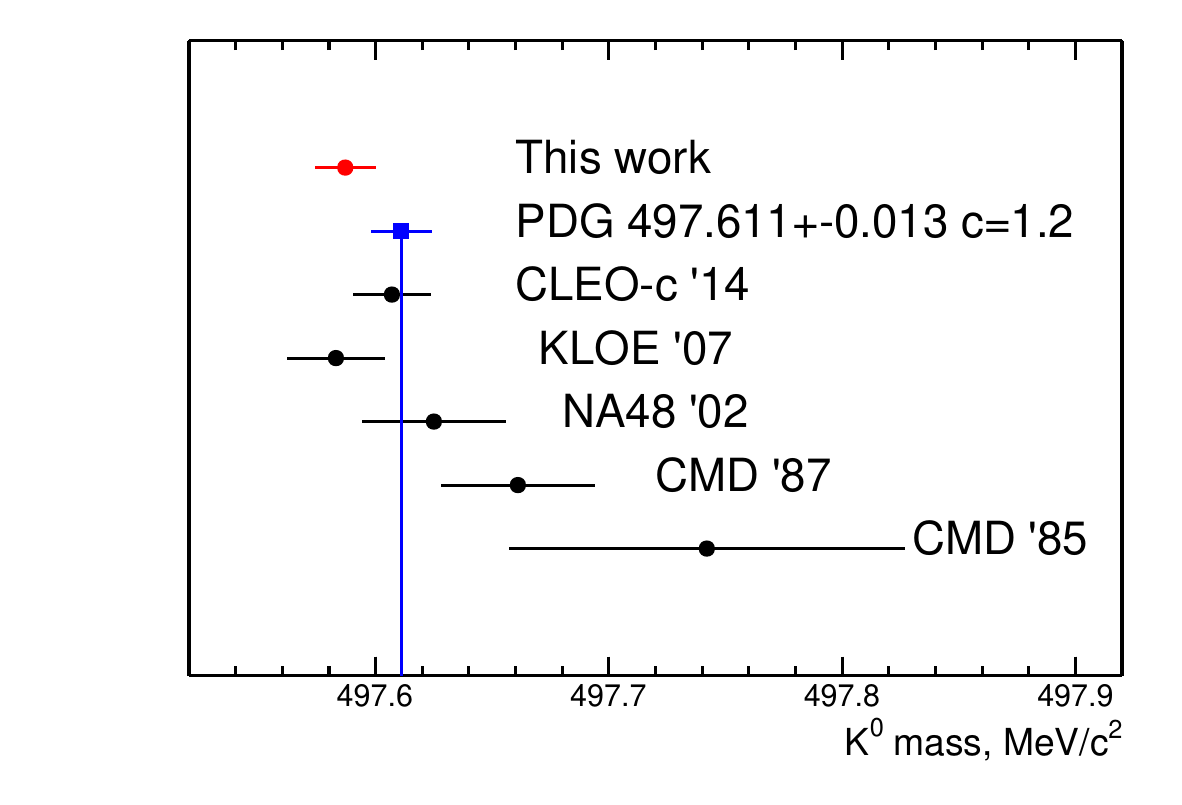}
\vspace{-0.3cm}
\caption{
\Kn~mass measurements in comparison with the CMD-3 result.    
}
\label{pdgcomp}
\end{center}
\end{figure}
\section{\boldmath Conclusion}
\label{conclusion}
\hspace*{\parindent}
Table~\ref{tabsyst} summarises the systematic uncertainties discussed
in Sec.~\ref{systematic}. The largest uncertainty comes from the beam energy
calibration using the world average $\phi$-meson mass. Two major
corrections, -0.030$\div$0.034\,\mevcc~for the beam energy spread and
0.015$\div$0.100\,\mevcc\\for the radiative effects, are applied with
 well under control systematic uncertainties.  
The \Ks~mass differs from the \Kn~ mass by a negligible value, therefore
our result is:
$$
m(\Kn) = 497.587 \pm 0.004(stat.) \pm 0.008(syst.) \pm 0.009(calibr.)~\mevcc,
$$
where the first uncertainty is statistical, the second is systematic,
and the third is the uncertainty from the energy calibration.
Assuming no
correlations between uncertainty sources, the overall 
uncertainty in our measurement is estimated as 0.0124 \mevcc. 
Figure~\ref{pdgcomp} shows our result in comparison with the other \Kn~
mass measurements.
Our result is in good agreement with all high statistical experiments
and has better accuracy.
\section*{\boldmath Acknowledgement}
\hspace*{\parindent}
The authors are grateful to the SND team
for their support of the energy measurement system and useful cooperation.
We thank the VEPP-2000 team for the excellent machine operation.

\begin{table}[tbh]

\caption{Beam energy, \Ecm energy spread, \Eb correction, edge angle, radiative correction to edge angle, and  
\Ks ~mass. Only statistical errors are shown.}
\label{tabular}
%\smallskip
\begin{center}
\begin{tabular}{|cccccc|}
\hline
\hline
\Eb , MeV&$\delta\Ecm$, keV&$\Delta E_{\phi}$, MeV&$\psi_c$, mrad
&$\Delta\psi_{rad}$, mrad&m(\Ks), MeV/c$^{2}$  \\
\hline
  500.954$\pm$0.015&   335$\pm$8&     0.0063$\pm$0.0003&   2862.01$\pm$1.87&   0.09$\pm$0.05&   497.602$\pm$0.047\\
  502.993$\pm$0.011&   356$\pm$11&   0.0092$\pm$0.0006&   2787.99$\pm$0.79&   0.42$\pm$0.05&   497.585$\pm$0.026\\
  504.798$\pm$0.008&   352$\pm$13&   0.0120$\pm$0.0009&   2734.55$\pm$0.90&   0.52$\pm$0.06&   497.59$\pm$0.0326\\
  507.868$\pm$0.009&   385$\pm$13&   0.0300$\pm$0.0026&   2657.06$\pm$0.27&   0.31$\pm$0.06&   497.589$\pm$0.014\\
  508.400$\pm$0.017&   351$\pm$20&   0.0283$\pm$0.0032&   2644.90$\pm$0.13&   0.24$\pm$0.06&   497.582$\pm$0.017\\
  508.957$\pm$0.016&   352$\pm$18&   0.0277$\pm$0.0029&   2633.03$\pm$0.11&   0.34$\pm$0.07&   497.591$\pm$0.016\\
  509.539$\pm$0.008&   373$\pm$8&     0.0112$\pm$0.0005&   2620.73$\pm$0.07&   0.41$\pm$0.06&   497.577$\pm$0.008\\
  509.970$\pm$0.008&   397$\pm$11&   -0.0160$\pm$0.0009&   2612.93$\pm$0.09&   0.69$\pm$0.07&   497.591$\pm$0.009\\
  510.454$\pm$0.007&   396$\pm$13&   -0.0343$\pm$0.0023&   2603.79$\pm$0.13&   1.18$\pm$0.07&   497.586$\pm$0.009\\
  511.046$\pm$0.007&   363$\pm$11&   -0.0303$\pm$0.0019&   2592.51$\pm$0.20&   1.83$\pm$0.07&   497.585$\pm$0.012\\
  511.466$\pm$0.014&   369$\pm$23&   -0.0285$\pm$0.0036&   2584.85$\pm$0.24&   2.07$\pm$0.08&   497.603$\pm$0.018\\
  513.868$\pm$0.012&   373$\pm$17&   -0.0158$\pm$0.0015&   2539.37$\pm$0.51&   1.73$\pm$0.09&   497.583$\pm$0.029\\
  516.908$\pm$0.018&   366$\pm$16&   -0.0091$\pm$0.0008&   2488.32$\pm$0.82&   1.35$\pm$0.10&   497.624$\pm$0.050\\
  519.894$\pm$0.018&   423$\pm$23&   -0.0087$\pm$0.0010&   2442.94$\pm$1.76&   1.05$\pm$0.23&   497.685$\pm$0.111\\
  524.902$\pm$0.023&   427$\pm$23&   -0.0060$\pm$0.0007&   2370.26$\pm$1.62&   0.81$\pm$0.12&   497.501$\pm$0.114\\
  \hline
\end{tabular}
\end{center}
\end{table}

\begin{table}[tbh]
  \caption{Summary of the systematic uncertainties study for the \Ks~
    mass measurement.}
\label{tabsyst}
%\smallskip
\begin{center}
\begin{tabular}{|c|c|c|c|}
\hline
\hline
 &Uncertainty source & Mass correction, \mevcc& Mass uncertainty
                                          \mevcc\\
\hline
  1 & Energy calibration &  0.0085 & 0.009\\
  \hline
  2 & Energy fit variation & - & 0.004\\
  3 & Energy spread at $\phi$ &  -0.030$\div$0.034 & 0.002\\
  4 & Y corrections & - & 0.001\\
  5 & |L$_{\rm LogY}$|<0.2 cut & - & 0.001\\
  6 & Radcorr (MC resolution) & 0.005$\div$0.010 & 0.001\\
  7 & Radcorr  (\Eb~ shift) & 0.002 & 0.002\\
  8 & Radcorr (strong $\psi_c$ fit) & 0.015$\div$0.100 & 0.003\\
  9 & Angle corrections&  - & 0.002\\
  10& Strong polar angle cut  $\pm$ 0.2 & - & 0.004\\
  \hline
   & Sum in quadrature  & & 0.012\\
 \hline
\end{tabular}
  \end{center}
\end{table}

\end{document}